
%

\def\NI{\noindent}
\long\def\UN#1{$\underline{{\vphantom{\hbox{#1}}}\smash{\hbox{#1}}}$}
\magnification=\magstep 1
\overfullrule=0pt
\hfuzz=16pt
\voffset=0.0 true in
\vsize=8.8 true in
   \def\NP{\vfil\eject}
   \baselineskip 20pt
   \parskip 6pt
   \hoffset=0.1 true in
   \hsize=6.3 true in
\nopagenumbers
\pageno=1
\footline={\hfil -- {\folio} -- \hfil}
\headline={\ifnum\pageno=1 \hfill October 1992 \fi}

\hphantom{AA}

\hphantom{AA}

\centerline{\UN{\bf Collective Effects in Random Sequential}}

\centerline{\UN{\bf Adsorption of Diffusing Hard Squares}}

\vskip 0.4in

\centerline{\bf Jian--Sheng Wang}
\centerline{\sl Department of Physics, Hong Kong Baptist College,
224 Waterloo Road, Kowloon, Hong Kong}

\centerline{\bf Peter Nielaba$^{\bf *}$}
\centerline{\sl Institut f\"ur Physik, Universit\"at Mainz,
Staudingerweg 7, D--6500 Mainz, Germany}

\centerline{\bf Vladimir Privman}
\centerline{\sl Department of Physics, Clarkson University,
Potsdam, New York 13699--5820, USA}

\vskip 0.6in

\centerline{\bf ABSTRACT}

We study by Monte Carlo computer simulations random sequential
adsorption (RSA) with diffusional relaxation, of lattice
hard squares in two dimensions. While for RSA without diffusion the
coverage approaches its maximum jamming  value (large-time
fractional coverage) exponentially, added diffusion allows the
deposition process to proceed to the full coverage. The approach to
the full coverage is consistent with the $\sim t^{-1/2}$ power law
reminiscent of the equilibrium cluster coarsening in models with
nonconserved order-parameter dynamics.

\vskip 0.3in

\NI {\bf PACS numbers:}$\;$ 68.10.Jy, 02.50.+s, 82.65.-i

\NP

Random sequential adsorption (RSA) models have been studied
extensively due to their relevance to deposition processes on
surfaces [1]. The depositing particles are represented
by hard-core extended objects; they are not allowed to overlap.
In monolayer deposition of colloidal particles and macromolecules
[2] one can further assume that the adhesion process is
irreversible. Once a particle is in place its relaxation on the
surface proceeds on time scales much larger than the deposition
process.

However, recent experiments on protein adhesion at surfaces
[3] indicate that in these systems effects of surface relaxation,
presumably due to diffusional rearrangement of particles, are
observable on time scales of the deposition process. The
resulting large-time coverage is denser than in fully
irreversible RSA and in fact it is experimentally comparable to
the fully packed (i.e., locally semi-crystalline) particle
arrangement. Studies of RSA with diffusional relaxation by
analytical means encounter several difficulties associated with
possible collective effects in hard-core particle systems at high
densities (such as, for instance, phase separation), and with the
possibility, in certain lattice models, of locally ``gridlocked''
vacant sites. The latter effect may actually prevent full coverage
in some models; this matter remains an open problem at this time.

Both difficulties are not present in $1D$: there are no
equilibrium phase transitions, traces of which might manifest
themselves as collective effects in $D>1$ deposition with
diffusion, and furthermore diffusional relaxation leads to simple
hopping-diffusion interpretation of the motion of vacant sites in
$1D$ which recombine to form larger open voids accessible to
deposition attempts. Thus, both extensive numerical studies and
their analytical interpretation were possible in $1D$\ \ [4]. For
higher-$D$ models however, no results were reported in the
literature, to our knowledge.

In this work we report extensive numerical simulations of the
RSA process with diffusional relaxation, for the lattice
hard-square model [5], i.e., the square-lattice hard-core model with
nearest-neighbor exclusion. This model is well studied for its
equilibrium phase transition [5] which is second-order with
disordered phase at low densities and two coexisting ordered
phases, corresponding to two different sublattice particle coverage
arrangements, at high densities. Another simplifying feature of
the hard-square model is that the only possible gridlocked
(locally frozen) vacancies are parts of domain walls (see further
below). As a result the coverage reaches the full crystalline limit
at large times, by a process of diffusional domain wall motion
leading to cluster growth reminiscent of quenched binary alloys
and fluids at low temperatures [6].

The approach to the full coverage is numerically consistent with
the $\sim t^{-1/2}$ law which in turn can be related to the domain
size growth with time as $t^{1/2}$ as expected for order-parameter
nonconserving dynamics. In the remainder of this work we first
report computational details and numerical results.  We then
describe the domain-growth interpretation of the dynamics at high
densities, as well as discuss some other collective effects
observed in our simulations.

In each Monte Carlo trial of our simulation on a $L\times L$ square
lattice with periodic boundary conditions, a site is chosen at
random. Then with probability $r$ we attempt to deposit a particle
and with probability $(1-r)$ we try diffusion.  In the case of
deposition, we check if the chosen site and its four
nearest-neighbor sites are all empty.  If indeed they are empty the
deposition is performed. The chosen site is marked as occupied.  If
however any of the five sites are already occupied, then the
deposition attempt is rejected and the configuration remains
unchanged.  In the case of diffusion, which is of course possible
only if there is a particle at the selected site, we choose at
random with equal probability a direction (up, down, left, or
right) and try to move this particle by one lattice spacing.  A move
is made if the targeted new site and its three nearest
neighbors different from the ``source'' site, are all empty.  If the
attempted move is not possible, the particle stays at its original
position.

A unit Monte Carlo time step is defined such that each lattice site
is checked once on average.  This corresponds to $L^2$ trials as
described earlier. This time scale, $T$, is conveniently related
to the physically more interesting time $t$ defined to have fixed
deposition attempt rate per site, with varying relative diffusion
attempt rate proportional to $(1-r)/r$,
$$ t=rT   \;   .  \eqno(1) $$

We implement the dynamics in two different ways.  The first one is a
straightforward simulation of the model.  However, it is
inefficient at late stages, when most trials are rejected.
Thus we have implemented the same dynamics with an event-driven
method [7]. We keep a list of all the possible moves, deposition
and diffusion.  Then we pick a move according to proper
probability and always carry it out.  The list is then updated if
necessary.  The Monte Carlo time is incremented according to
$\Delta T = - (\ln x)/R$, where $x$ is a uniform random number
between 0 and~1, while $R$ is the rate of the system
configuration changes in the original dynamics.

Suppose our list contains $N_{\rm depo}$ sites available for
deposition and $N_{\rm diff}$ possible diffusional moves. Each
occupied lattice site contributes $0,1,2,3$ or 4 possible
diffusional moves, while each empty lattice site contributes 0 or 1
deposition counts, all depending on the nearest-neighbor site
configuration. Then $R$ is calculated as
$$ R = r N_{\rm depo} + {1\over 4} (1-r)N_{\rm diff}\; . \eqno(2) $$

Numerical estimates were obtained for the following
quantities. The coverage, $\theta(T)$, was defined as the total
number of sites occupied divided by $L^2/2$. The deposition process
always began with empty substrate so that the coverage $\theta$
increased from 0 to 1 at full saturation. The ``susceptibility''
measuring fluctuations of the magnetization, was defined by
$$\chi = L^2 \bigl[ \langle m^2 \rangle -
\langle |m| \rangle^2 \bigr] \; , \eqno(3) $$
where the average $\langle\ \rangle$ is over independent runs.
The magnetization or order parameter was defined as usual [5]
by assigning ``spin'' values $+1$ to particles on one of the
sublattices and $-1$ on another sublattice. Empty sites on both
sublattices were not counted (effectively having spin values 0).
Thus the magnetization magnitude $|m|$
is defined as the difference of the number of particles deposited on
two sublattices, normalized by the lattice size $L^2$. The values
of $|m|$ are thus from 0 to $1\over 2$.

The effective domain size, $\ell (T)$, was defined as in
equilibrium-model studies of cluster coarsening [6], by
$$\ell  = 2L \sqrt{\langle m^2\rangle} \; . \eqno(4) $$
Here the normalization is such that a uniform state
(single-domain) gives the size of the system, $L$.

Series of snapshots of the coverage buildup are shown in Fig.~1.
As is usually done for the equilibrium hard-square system
[5], particles are represented by squares of size $\sqrt{2}
\times \sqrt{2}$, rotated 45$^\circ$ with respect to the
original square lattice on which the particle centers are
deposited.  Particles on the even and odd sublattices (the sum of
the $x$ and $y$ coordinates even or odd) are shown in different
shades.

The time-dependence of the coverage is illustrated in Fig.~2.
The general features are similar to those found in the $1D$
studies [4]. For fixed deposition rate, corresponding to the
time scale $t$ defined in (1), added diffusion [rate
$\sim (1-r)/r$] always speeds up coverage growth. For high
coverage, the $r<1$ plots are reminiscent of domain growth in
phase-separation models [6].

This similarity with domain growth dynamics can be made more
quantitative.  Let us first consider the coverage $\theta(T)$ for
large times.  In the case of deposition rate $r=1$ (no diffusion),
the approach to the jamming coverage $\theta (\infty) \simeq
0.728 < 1$\ \ [8] is
generally exponentially fast for lattice models [9].  With
diffusion, one can always reach the full coverage
$\theta( \infty ) = 1$.  However, the approach to the full
coverage is slow, power-law, as indicated in Fig.~2. In $1D$, the
power-law behavior was related to the coverage growth at large
times
by the process of hopping and recombination (opening up deposition
sites) of small empty regions [4].

The coverage growth mechanism for large times, in the $2D$
hard-square model is instead due to interfacial dynamics. As
illustrated in Fig.~1, the void space at late times consists of
domain walls separating spin-up and spin-down ordered regions.
Since a typical domain has area $\sim \ell^2 (T)$ and boundary
$\sim \ell(T)$, we anticipate that for large times
$$ 1 - \theta (T) \propto \ell^{-1} (T) \; . \eqno(5) $$
Indeed, $1-\theta$ is just the void area fraction. The
large-time behavior of the domain size in models with nonconserved
order-parameter dynamics is typically diffusional $\sim T^{1/2}$.
For $\ell (T)$ we report the direct numerical verification later.
For the coverage, we found that the data roughly fit the power law,
$$ 1- \theta(T) \propto T^{-1/2} \; , \eqno(6) $$
for $T > 10^3$ in typical runs such as illustrated in
Fig.~2. Thus, the RSA quantity $1-\theta(T)$ behaves analogously to
the energy excess in equilibrium domain growth problems.

However, on careful examination we noted that the slopes obtained
by least-square fits of the straight-line portions of the curves
for large $T$ were not exactly $-0.5$ but rather ranged from about
$-0.48$ to about $-0.52$ when the rate parameter $r$ was
decreased from $0.8$ to $0.01$, corresponding to accelerating
the relative diffusion rate. These differences could not be
fully attributed to statistical errors. Presumably, they represent
effects of corrections to the leading power law behavior.

The ``susceptibility'' $\chi$ for a given finite size $L$ has a
peak and then decreases to zero, indicating long-range order for
large $T$; see Fig.~3.  The peak location seems size-dependent, at
$T_{\rm peak} \propto L^2$; thus, it is difficult to observe for
large system sizes.  Since finite-size effects set in for $\ell (T)
\sim L$, which given the ``bulk'' power law $\ell (T) \sim T^{1/2}$
leads precisely to the criterion $T \sim L^2$, we expect this
maximum in fluctuations to be a manifestation of the ordering
process at high densities.

For equilibrium hard squares [5], the critical-point density
corresponds to coverage $\theta \simeq 0.742$. However, in all our
simulations, even with the fastest relative diffusion rate (the
case $r=0.01$), we found no interesting features in $\chi (T)$ for
times for which the coverage was near the critical value. Since
numerical effort to reach a given coverage increases with
increasing the diffusion rate at the expense of deposition
attempts, there still remains a numerical challenge to observe the
buildup of rounded critical-point fluctuations in $\chi$ or other
quantity, for RSA with diffusion. Our present simulations turn out
to be sensitive only to those collective effects which
are associated with ordering at higher than critical-point
coverages.

In Fig.~4, the effective domain size $\ell(T)$ is plotted vs.~time
for $r=0.1$.  The large-time asymptotic law is well
fitted by $T^{1/2}$. As already mentioned, the late-stage dynamics
of our model is similar to the kinetics of ordering by
quenching an equilibrium system into a two-phase region.  In
fact, our model can be mapped approximately to the Ising
model at zero temperature.  Note that the straight-line, diagonal
sections of the domain boundaries separating two phases, can not
move.  Only the corners can evolve due to diffusion, just like
corners of interfaces
in the Ising models at zero temperature evolve by
order-parameter nonconserving dynamics.  It has been known that
the $t^{1/2}$ law is robust, independent of the number of
ordered phases, dimensionality, and details of interactions in
domain growth problems [6].

In summary, we found that when diffusion is introduced in random
sequential adsorption processes, the system can relax to the full
coverage, consistent with experiments on protein adsorption at
surfaces.  The approach to the full coverage is slow (power
law).  The dynamics for large coverages  is governed by domain
growth.

This research was supported in part by the Hong Kong Baptist
College Faculty Research Grant and by the
DFG-Sonderforschungsbereich 262/D1 (Germany).

\NP

\centerline{\bf REFERENCES}

{\frenchspacing

\item{${}^{\bf *}$} DFG Heisenberg Fellow

\item{[1]} For recent reviews see M.C.~Bartelt and V.~Privman, Int.
J.~Mod. Phys. B{\bf 5}, 2883 (1991); J.W.~Evans, Rev. Mod. Phys.
(1992), in print.

\item{[2]} J.~Feder and I.~Giaever, J.~Colloid Interface Sci. {\bf
78}, 144 (1980); G.Y.~Onoda and E.G.~Liniger, Phys. Rev. A{\bf
33}, 715 (1986); N.~Ryde, N.~Kallay and E.~Matijevi\'c, J.~Chem.
Soc. Faraday Trans. {\bf 87}, 1377 (1991).

\item{[3]} J.J.~Ramsden, preprint (1992); see also J.~Phys. Chem.
{\bf 96}, 3388 (1992).

\item{[4]} V.~Privman and P.~Nielaba, Europhys. Lett. {\bf 18}, 673
(1992); P.~Nielaba and V.~Privman, Mod. Phys. Lett. B {\bf 6},
533 (1992); V.~Privman and M.~Barma, J.~Chem. Phys. (1992), in
print.

\item{[5]} L.K.~Runnels, in {\sl Phase Transitions and Critical
Phenomena}, Vol.~2, p.~305, C.~Domb and M.S.~Green, eds.
(Academic, London, 1972).

\item{[6]} J.D.~Gunton, M.~San~Miguel, P.S.~Sahni,
{\sl Phase Transitions and Critical
Phenomena}, Vol.~8, p.~267, C.~Domb and J.L.~Lebowitz, eds.
(Academic, London, 1983); O.G.~Mouritsen, in {\sl Kinetics and
Ordering and Growth at Surfaces}, p.~1, M.G. Lagally, ed. (Plenum,
NY, 1990); A.~Sadiq and K.~Binder, J.~Stat. Phys. {\bf 35}, 517
(1984).

\item{[7]} K.~Binder, in {\sl Monte Carlo Methods in Statistical
Physics}, 2nd. ed., p.~1, K.~Binder, ed. (Springer-Verleg,
Berlin, 1986).

\item{[8]} Numerical studies of the RSA process of
hard squares without diffusion were reported by P.~Meakin,
J.L.~Cardy, E.~Loh and D.J.~Scalapino, J.~Chem. Phys. {\bf 86}, 2380
(1987); R.~Dickman, J.--S.~Wang and I.~Jensen, J.~Chem. Phys. {\bf
94}, 8252 (1991); see also references quoted therein.

\item{[9]} V.~Privman, J.--S.~Wang and P.~Nielaba, Phys. Rev. B{\bf
43}, 3366 (1991).

}

\NP

\centerline{\bf FIGURE CAPTIONS}

\

\noindent\hang{\bf Fig.~1.}\ \
Configurations of the hard-square RSA-with-diffusion model
obtained
with the deposition rate parameter $r=0.1$, at times $T=10$, 20, 30,
and 100.  Particles centered on the even sublattice are shown in
grey,
while those on the odd sublattice are shown in black. Lattice size
was $32\times 32$.

\noindent\hang{\bf Fig.~2.}\ \
Void area fraction, $1-\theta$,
plotted against time $t=rT$ on a double-logarithmic scale, for the
deposition rate parameter $r=1$, $0.8$, $0.1$, $0.01$, and
system size 521$\times$521. The data were averaged over 800,
600,  800, and 400  runs for $r=1$, 0.8, 0.1, and 0.01,
respectively.  The dashed-dotted line has slope of $-{1\over 2}$.

\noindent\hang{\bf Fig.~3.}\ \
Fluctuation of the magnetization, $\chi$, vs.~time $T$, for system
sizes $L=$4, 8, 16, 32, and 64.  The deposition rate was $r=0.1$.
The data were averaged over $10^5$ to $10^6$ runs.

\noindent\hang{\bf Fig.~4.}\ \
Effective domain size $\ell$, plotted vs.~time $T$
on a double-logarithmic scale, with $r=0.1$ and
system size 256$\times$256.  The data were averaged over 1500
runs.

\bye